\documentclass[prd,aps,showpacs,preprintnumbers,amssymb]{revtex4}
\usepackage{graphicx}
\usepackage{dcolumn}
\usepackage{bm}
\begin{document}
\title{QED Plasma at Finite Temperature up to  Two Loops }

 \author{Samina S. {\sc Masood}}
 \email{masood@uhcl.edu}

 \affiliation{Department of Physics, University of
Houston Clear Lake, Houston TX 77058}

\date{August 2018}

\begin{abstract}
 We study the vacuum polarization tensor of QED (quantum electrodynamics) at high temperatures up to the two loop levels and its effect on the electromagnetic properties of a medium. One loop corrections to QED coupling vanish at low temperatures (T$\leq 10^{10}K$), but they play an important role at high temperature ( T$\geq 10^{10}$ K) to study the behavior of QED medium at these temperatures. At low temperatures ( $T \leq m_e$)higher order loops give a tiny correction due to the coupling of radiation with matter and an overlap of hot photon loop with cold fermion loop contributes to this effect. These higher loop contributions does not affect the convergence of perturbative series, and renormalizability of QED is guaranteed at temperatures around neutrino decoupling. We use the renormalization scheme of QED at finite temperature in real-time formalism to study the dynamically generated mass of photon indicating the plasmon production in such a medium. Temperature dependence of this QED plasma parameters is discussed.  We explicitly show that this behavior of a thermal medium exists upto temperatures of a few MeV  only. We compare the first order and second order effects upto the 4MeV temperature and demonstrate that the higher order contributions are smaller than the lower order contributions proving the renormalizability of the theory. The lowest order contributions are sufficiently smaller than the original value as well. 
\end{abstract}

\maketitle


\section{INTRODUCTION}

In quantum field theory, thermal background effects are incorporated through the radiative corrections. Renormalization of gauge theories [1-2] at finite temperature [3-11] requires the renormalization of gauge parameters of the corresponding theory. The propagation of particles and the electromagnetic properties of media are also known to modify in the framework of real-time formalism [7-9]. Masses of particles are shown to increase with temperature at the one-loop level [11-22], the two-loop level [23-26] and presumably to all loop levels [27]. At the higher-loop level, the loop integrals have a combination of cold and hot terms which appear due to the overlapping propagator terms in the matrix element. Higher loop calculations are too cumbersome to be performed analytically using the perturbation theory. Order by order cancellations of singularities [12] cannot be shown in the imaginary time formalism. Therefore, the effective potential approach at finite temperature [28] is used to study the overall effects. However, renormalizability can only be tested in detail when order-by-order cancellation of singularities [29] is shown in real-time formalism. 

\bigskip
However, the gauge bosons acquire dynamically generated mass due to plasma screening effect [16-19], at the one-loop level. It helps to determine the changes in electromagnetic properties of a hot medium. In hot gauge theories $m_{e}$ is the electron mass and corresponds to $10^{10}$ K. The vacuum polarization tensor in order $\alpha$ (the QED coupling parameter) does not acquire any hot corrections from hot photons in the heat bath [16] because of the absence of self-interaction of photons in QED. The photon can only interact with the medium in the presence of electrons, which decouple at high temperatures ($\geq 1MeV or 10^{10} K$). Some of the QED parameters have already been calculated for such systems. The effective value of the electric charge increases due to the interaction of charge in the medium and consequently leads to modifications in electromagnetic properties of a medium due to the enhancement in QED coupling. These type of calculations are discussed up to the two loop level. These QED couplings at high density can change an ordinary fluid in a relativistic plasma and give a nonzero value of dynamically generated mass [16-19], which can be treated as a plasma screening mass and causes Debye shielding. We calculate the Debye shielding length of QED plasma as a function of temperature for a QED system which can be identified as QED plasma.
\bigskip

In this paper, we use the renormalization scheme of QED in real-time formalism to calculate the parameters of QED plasma in terms of thermally corrected renormalized values of electric charge, mass and wavefunction of electron [29]. These thermal contributions to the fundamental parameters of the theory help to determine the effective parameters of the theory under given statistical conditions.  In the real-time formalism, all the calculations are done in the rest frame of the heat-bath to re-establish covariance at the cost of Lorentz invariance [7, 20]. The particle propagators include temperature dependent (hot) term in addition to the temperature independent (cold) term [5]. At the higher loop level, the loop integrals involve an overlap of hot and cold terms in particle propagators. This makes the situation cumbersome and getting rid of these singularities needs much more involved calculations. Sometimes $\delta (0)$ type pinch singularities appear in Minkowski space. Cancellation of singularities is not  obvious, and this entire scheme works perfectly fine for $T<4m_e$ up to the two loop level [26-28]. 

\bigskip

In the next section, we discuss the vacuum polarization tensor in different ranges of temperature up to the second order in $\alpha$.  Section III is comprised of the calculation of the plasma parameters up to the two-loop levels. The plasma screening mass and plasma frequencies up to the second order in alpha are discussed in section IV. We conclude this paper by the discussions of application of these results to certain physical systems in section V.

\section{VACUUM POLARIZATION TENSOR IN QED}

The lowest order vacuum polarization diagram of QED is given in Fig. 1 and is essential to prove the renormalizability of QED at the lowest level of perturbation theory. This diagram mainly contributes to the QED coupling, which is not affected at low energies. 

\begin{figure}[!hth]
\begin{center}
  \begin{tabular}{c}
    \mbox{\includegraphics[width=2.0 in,angle=0]
{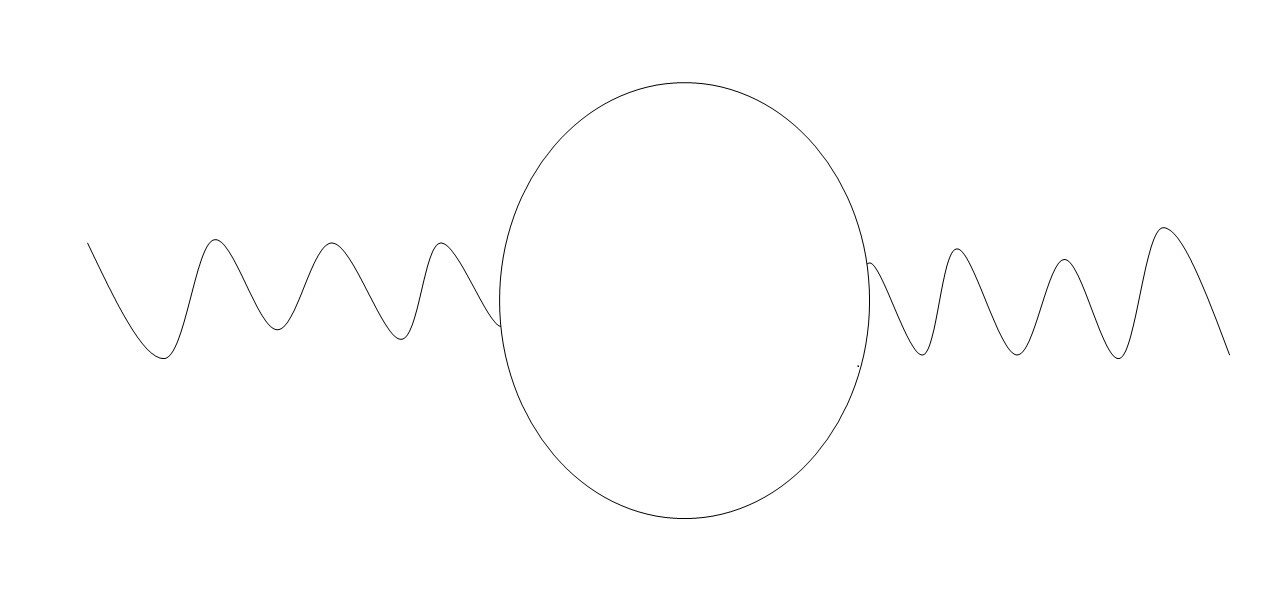}}
  \end{tabular}
\end{center}
\caption{First order vacuum polarization diagram in QED: Photon produces a virtual electron loop during propagation which can couple with the electrons of the medium.} 
\end{figure}

The vacuum polarization tensor of QED as a function of photon momentum p and the loop momentum k is expressed as:  
\begin{eqnarray}
\Pi _{\mu \nu }(p) &=&e^{2}\int \frac{d^{4}k}{(2\pi )^{4}} \gamma _{\mu }S_{\beta }(k)\gamma _{\nu}S_{\beta }(k+p)  
\end{eqnarray} 

and the fermion propagator is replaced by the thermally corrected propagator to incorporate thermal effects. The thermal propagator of QED is calculated using the fermion propagator, 
\begin{eqnarray}
S_{\beta}(k) = \not k [\frac{1}{k^2+{m_e}^2}+2\pi n_F(k)] \nonumber
\end{eqnarray} 
with the fermion distribution function,
\begin{eqnarray}
n_F(k)=\frac {1}{e^{\beta k}+1} \nonumber
\end{eqnarray} 

\bigskip
in usual notations. The absence of a photon propagator in Eq.(1) corresponds to the lack of hot electrons in the propagating medium causing the lack of thermal contribution at low temperature. Whereas, at the two loop level, there is an overlap of electron and photon propagators  and the electron loops in vacuum can pick up thermal corrections from hot photons in the background through cold electron loops as well. These small contributions do not exceed the expansion parameter of the perturbation series before the decoupling of neutrinos take place. In the presence of neutrinos, after decoupling, they can interact with radiation [30-32] and the relevant electroweak processes cannot be ignored.  

\bigskip 
The vacuum polarization tensor of photon at the two-loop level gives the second order hot corrections to charge renormalization constant of QED at low temperature. Second order corrections to the vacuum polarization are mainly described by three diagrams given in Figure 2. These contributions are calculated from QED vertex type corrections Fig. (2a) and the electron self mass type corrections of Fig. (2b) inside the vacuum polarization tensor. In Fig. (2c), the mass counter term has to be included in the vacuum polarization to cancel singularities. 

\begin{figure}[!hth]
\begin{center}
  \begin{tabular}{c}
    \mbox{\includegraphics[width=4in,angle=0]
{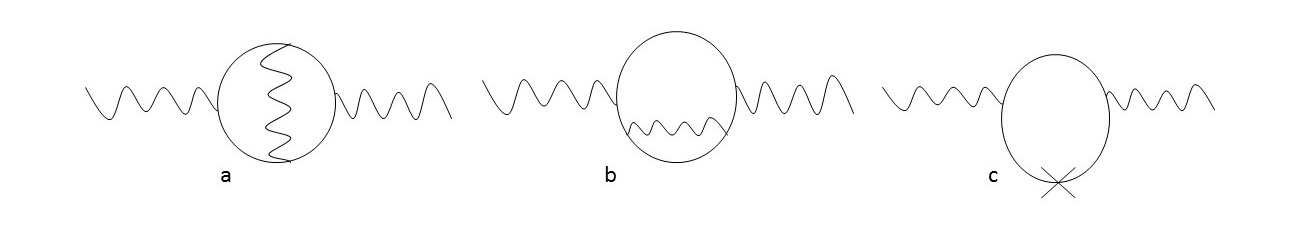}}
  \end{tabular}
\end{center}
\caption{Second order vacuum polarization diagrams in QED. It includes second order corrections due to : (a) QED vertex corrections , (b) electron selfmass and (c) is the counter-term.} 
\end{figure}

The diagrams in Fig.2 are used to calculate the second order thermal contributions to the vacuum polarization tensor due to the hot photon loops at low temperature, i.e., $T\ll m_{e}$ . In this scheme of calculations, the vacuum polarization tensor of photon in Fig. (2a) is given by

\begin{eqnarray}
\Pi _{\mu \nu }^{a}(p) &=&e^{4}\int \frac{d^{4}k}{(2\pi )^{4}}\int \frac{%
d^{4}l}{(2\pi )^{4}}tr[\gamma _{\mu }S_{\beta }(k)\gamma _{\rho
}D_{\beta}^{_{\rho _{\sigma }}}(l)S_{\beta }(k+l)  \nonumber \\
&&\gamma _{\nu }S_{\beta }(k+l-p)\gamma _{\rho }S_{\beta }(k-p)],
\end{eqnarray}

while  Fig.(2b) is

\begin{eqnarray}
\Pi _{\mu \nu }^{b}(p) &=&e^{4}\int \frac{d^{4}k}{(2\pi )^{4}}\int \frac{%
d^{4}l}{(2\pi )^{4}}tr[\gamma _{\mu }S_{\beta }(k)\gamma _{\rho
}D_{\beta}^{_{\rho _{\sigma }}}(l)  \nonumber \\
&&S_{\beta }(k+l)\gamma _{\sigma }S_{\beta }(k)\gamma _{\nu
}S_{\beta }(p-k)].\
\end{eqnarray}

where the photon propagator is given as, 

\begin{eqnarray}
D^{\mu\nu}_{\beta}(l)= [\frac{g^{\mu\nu}}{k^2+{m_e}^2}+2\pi n_B(l)] \nonumber
\end{eqnarray}
with the photon distribution function, 
\begin{eqnarray}
n_B(l)=\frac {1}{e^{\beta l}-1}. \nonumber
\end{eqnarray} 

\bigskip
Renormalization of QED is established by order-by-order cancellation of singularities [23-26]. Therefore the contributions of all of the same loop-order diagrams are required to be added together to demonstrate the cancellation of singularities. Due to the choice of the rest-frame of the heatbath,  Lorentz invariance breaks (to maintain gauge invariance) in real-time formalism.  For this purpose, we need to get rid of hot divergences appearing due to the presence of the hot loops without distributing them over the cold loops. The  energy integration has to be performed before 3- momenta integrals.  Once the energy integration is done, the three dimensional momentum integrals can even be performed using the techniques of covariant field theory in vacuum. 

\bigskip
Due to the imposed covariance in the propagators, physically measurable contributions to the couplings can be obtained in terms of longitudinal and transverse components of the polarization tensor through the contraction of vacuum polarization tensor $\Pi _{\mu \nu }$ with $g^{\mu\nu}$ in Minkowski space. As a result of this calculation, we obtain a nonzero value of the  longitudinal component ($\Pi_L$) due to the coupling of photon to the medium through the dynamically generated mass. 

\bigskip
Contraction of the 4-velocity of the heatbath $u^\mu$ with the polarization tensor $\Pi _{\mu \nu }$  gives the longitudinal and transverse components of this tensor while the photon is propagating through the medium and is calculated as:

\begin{equation}
u^{\mu }u^{\nu }\Pi _{\mu \nu }(p,T)=-\frac{2\alpha ^{2}T^{2}}{3}(1+\frac{p_{0}^{2}}{2m^{2}}),
\end{equation}

and

\begin{equation}
g^{\mu \nu }\Pi _{\mu \nu }(p,T)=\frac{\alpha ^{2}T^{2}}{3}.
\end{equation}
Moreover, the longitudinal and the transverse components of the vacuum polarization tensor that can be calculated from Eqs. (2) and (3) are:

\begin{eqnarray}
\Pi _{L}(p,T) &=&-\frac{p^{2}}{|\mathbf{p}|^{2}}u^{\mu }u^{\nu }\Pi
_{\mu
\nu }(p,T)  \nonumber \\
&=&\frac{2\alpha ^{2}T^{2}p^{2}}{3|\mathbf{p}|^{2}}(1+\frac{p_{0}^{2}}{2m^{2}}),
\end{eqnarray}
and
\bigskip
\begin{eqnarray}
\Pi _{T}(p,T) &=&-\frac{1}{2}[\Pi _{L}(p,T)-g^{\mu \nu }\Pi _{\mu
\nu}^{b}(p,T)]  \nonumber \\
&=&\frac{\alpha ^{2}T^{2}}{3}[\frac{1}{2}-\frac{p^{2}}{|\mathbf{p}|^{2}}(1+\frac{p_{0}^{2}}{2m^{2}})].
\end{eqnarray}
These components of the vacuum polarization tensor are used to determine the electromagnetic properties of a medium with hot photons and electrons. Light is transverse in nature in vacuum and even in a medium at normal temperatures and densities.  However, the nonzero value of the longitudinal component in Eq.(6) indicates that a temperature dependent longitudinal component arises due to the self-interaction of photons at high temperature, which is greater than $10^9$ K  and may exist in the early universe or inside stars.  This longitudinal component indicates the coupling of a  photon with the medium through virtual photons and gives rise to a dynamically generated mass of photon. This dynamically generated mass can be referred to as the plasma shielding mass and photon may be treated as a plasmon due to this distinct behavior. This behavior is observed at high densities [18, 33, 34] at the one loop level as well but we ignore it at the two loop level here.

\section{ELECTROMAGNETIC PROPERTIES OF A MEDIUM}

The QED coupling becomes temperature dependent due to the dynamically generated temperature dependent mass of photon. Since the photon couples with the medium through vacuum polarization, it picks up thermal corrections from the higher order coupling of radiation (photon) with matter (electron) in the medium. As a result of this interaction, QED coupling start to increase with the rise of temperature. 
\subsection{Low temeprature properties}
One loop corrections are only possible when $T\geq 10^{10}K$ which is enough energy for a photon to produce an electron-positron pair during its propagation and let the virtual electrons couple with the medium. Using the standard method of evaluation of charge renormalization constant of QED [26], the electron charge renormalization constant can be expressed as a quadratic function of temperature in the units of electron mass.
\begin{equation}
Z_{3}=1+\frac{\alpha ^{2}T}{6m^{2}}^{2}.
\end{equation}
Absence of the first order term  in $\alpha$ corresponds to the absence of the photon loops [16, 23, 29] in the first order diagram in Fig.1(as expected). The corresponding value of the QED coupling constant comes out to be
\begin{equation}
\alpha _{R}=\alpha (T=0)(1+\frac{\alpha ^{2}T^{2}}{6m^{2}}).
\end{equation}

The electromagnetic properties of media depend on the coupling of photon with the medium and are determined using techniques of quantum statistical mechanics. Then the parameters of the  electromagnetic theory become a function of statistical variables such as temperature and chemical potential. One-loop corrections are not modified at low temperatures, whereas the second order corrections minutely contribute at low temperature. Incorporating the thermal loops, longitudinal and transverse components of the wavenumber due to the plasma screening mass of photon show a  quadratic temperature dependence. For this purpose, we take the limit  $|\mathbf{p}|\longrightarrow 0$\ to obtain

\begin{equation}
\kappa _{L}^{2}\longrightarrow \lim_{|\mathbf{p}|\longrightarrow
0}\Pi _{L}(0,|\mathbf{p}|,T)=\frac{2\alpha ^{2}T^{2}}{3m^2},
\end{equation}

\begin{equation}
\kappa _{T}^{2}\longrightarrow \lim_{|\mathbf{p}|\longrightarrow
0}\Pi _{T}(0,|\mathbf{p}|,T)=\frac{\alpha ^{2}T^{2}}{2m^2}.
\end{equation}

On the other hand if we set $p_{0}=0$ with $|\mathbf{p}|\longrightarrow 0$ the thermal contribution to the photon frequencies are given as:

\begin{equation}
\omega _{L}^{2}\longrightarrow \lim_{|\mathbf{p}|\longrightarrow 0}\Pi _{L}(|
\mathbf{p}|,|\mathbf{p}|,T)=0,
\end{equation}

\begin{equation}
\omega _{T}^{2}\longrightarrow \lim_{|\mathbf{p}|\longrightarrow 0}\Pi _{T}(|
\mathbf{p}|,|\mathbf{p}|,T)=\frac{\alpha ^{2}T^{2}}{6m^2}.
\end{equation}

The difference between the longitudinal and transverse components of the vacuum polarization tensor in the limit $p^{2}\longrightarrow 0$, measures the dynamically generated mass of the photon. Electromagnetic properties of a medium change due to this radiatively generated effective mass.

\subsection{High temeprature properties}

At  very high temperatures when ($T\geq10^{10}$ K or $T\geq m_e$), the two-loop level contribution of the order of $T^2/m^2_e$ is given as
\smallskip
\begin{equation}
Z_{3}=1+\frac{2\alpha T^2}{3m^2} +\frac{\alpha ^{2}T^2}{6m^{2}}+O\left( \frac{T^4}{m^4}\right).
\end{equation}
\smallskip

and in the limit  $|\mathbf{p}|\longrightarrow 0$, we get

\begin{equation}
\kappa _{L}^{2}\longrightarrow \lim_{|\mathbf{p}|\longrightarrow
0}\Pi _{L}(0,|\mathbf{p}|,T)=\frac{4\alpha T^{2}}{3m^2} + \frac{2\alpha ^{2}T^{2}}{3m^2}+O\left( \frac{T^4}{m^4}\right).
\end{equation}

\begin{equation}
\kappa _{T}^{2}\longrightarrow \lim_{|\mathbf{p}|\longrightarrow
0}\Pi _{T}(0,|\mathbf{p}|,T)=\frac{\alpha^{2} T^{2}}{2m^2}+O\left( \frac{T^4}{m^4}\right).
\end{equation}

It can be clearly seen from Eq.(10) that there is no first order thermal contribution to $\kappa_T$ at low temperatures, whereas Eq. (15) has an $\alpha$ dependent term. On the other hand if we set $p_{0}=0$ with $|\mathbf{p}|\longrightarrow 0$, we obtain

\begin{equation}
\omega _{L}^{2}\longrightarrow \lim_{|\mathbf{p}|\longrightarrow 0}\Pi _{L}(|
\mathbf{p}|,|\mathbf{p}|,T)=0,
\end{equation}

\begin{equation}
\omega _{T}^{2}\longrightarrow \lim_{|\mathbf{p}|\longrightarrow 0}\Pi _{T}(|
\mathbf{p}|,|\mathbf{p}|,T)=\frac{\alpha T^{2}}{6m^2}+\frac{\alpha ^{2}T^{2}}{6m^2}+O\left( \frac{T^4}{m^4}\right).
\end{equation}

It  shows that at low energies, the longitudinal contributuion to frequencies vanishes at the first loop level. Therefore, the plasma parameters can be found from the longitudinal behavior of photon.  

\section{Debye Shielding in QED Plasma}

The plasma parameters in QED are calculated from $\kappa_L$ from equations (10) and (15) for the relevant ranges of temperature. Then
Debye length is inversely proportional to $\kappa_L$ such that

\begin{equation}
\lambda_D=\frac {1}{\kappa_L}
\end{equation}

giving the plasma screeninng length as,

\begin{equation}
\lambda_D=\sqrt{\frac {3}{2}} \frac{m}{\alpha T}
\end{equation}
for low temperature. This shows that the plasma length will decrease with the rise in temperature, if everything else remains constant. However, this expression will be modified for extremely high temperatures and Eq. (15) will be substituted for $\kappa_L$ giving,

\begin{equation}
\lambda_D=\left[\frac{4\alpha T^{2}}{3m^2} + \frac{2\alpha ^{2}T^{2}}{3m^2}+O(T^4)\right]^{-2}
\end{equation}
The dominent contribution to the Debye shielding is also due to the first order effect. The corresponding mass of the plasmon can be calculated from $k^2=\kappa^2-\omega^2$

\section{RESULTS\ AND DISCUSSIONS}

We have previously noticed [16] that the low temperature (T$\leq m_e$) effects on the vacuum polarization tensor vanish because of the absence of hot electron background. The selfmass of electron is thermally corrected due to the radiation background. However, we have found that thermal corrections to vacuum polarization tensor are nonzero at the higher loop level even at low temperatures because the cold electron loops can overlap with the hot radiation loops and pick up tiny thermal corrections.

\bigskip
It is  worth mentioning that the sizeable thermal corrections at high temperatures (T$\geq m_e$) are not always strong enough to create QED plasma. High temperatures indicate large kinetic energy and the interaction may still be weak enough to treat the system as an ideal gas. However, if the temperature is large and the volume is small, the mutual interaction between electrons is negligibly small as compared to thermal energy of particles. Therefore, a detailed study is required to understand the behavior of such systems which may have a plasma phase for a short time at some particular temperature. However, due to the exponential dependence on temperature, we can use high or low temperature limits for smaller changes due to the quadratic dependence in above equations (see Eqns.10-18). 

\bigskip
Thermal corrections, which lead to the modifications in electromagnetic properties of a medium itself are expected to give larger contribution at higher loop levels. However, most of these terms are finite and the order by order cancellation of singularities [12] is observed through the addition of all the same order diagrams. However, the renormalization of the theory can only be proved if covariant hot integrals are evaluated before the cold divergent integrals on the mass shell. Once the hot loop energies are integrated out, the usual vacuum techniques of Feynman parameterization and dimensional regularization can be applied to get rid of vacuum singularities. It is also worth-mentioning that all the hot corrections give a similar $T^{2}$ dependence. The incorrect order of integration gives an increased order of T due to the overlap with the vacuum divergences. This unusual behaviour of hot integrals induces additional temperature dependent divergences due to the overlap of hot and cold terms. Whereas, the usual regularization techniques of vacuum theory like dimensional regularization would only be valid in a covariant framework. Higher order terms may then violate renormalizability with the inverted order of integration.

\bigskip
 The presence of the statistical contribution of the photon propagator modifies the vacuum polarization and hence the electron charge which leads to changes (though small) in the electromagnetic properties of the hot medium even at low temperatures. Mass, wavefunction, and charge of electron are renormalized [29] in the presence of heat bath. These renormalized values give the dynamically generated mass of photon and its effective charge in such a background modifying dielectric constant and magnetic permeability of a hot medium[29]. The longitudinal component of vacuum polarization tensor $\Pi_L$ vanishes at low temperature as $p^{2}\longrightarrow 0$. However, the transverse component has an extremely small additional thermal contribution and  Eq. (7) reduces to, 

\begin{equation}
\Pi _{T}(p)=\frac{\alpha ^{2}T^{2}}{6},
\end{equation}

It is worth-mentioning that in the real-time formalism, the propagator has two additive terms, the vacuum term and the temperature dependent hot term. Therefore, in the second order perturbation theory we get a purely hot term ($~\alpha^2 T^4/m^4$), purely cold terms (~$T^0$) and the overlapping hot and cold terms ($~\alpha^2 T^2/m^2$). These terms can only be obtained in this formalism at the two-loop level. So, the overall result comes out to be a combination of all of these terms. For example, Eq.(9) can be written as, 

\begin{equation}
\alpha _{R}=\alpha (T=0)(1+ \frac{\alpha 
^{2}T^{2}}{6m^{2}}+O(\frac{T^{4}}{m^{4}})).
\end{equation}

We restrict ourselves in the temperature range where the last term ($~\alpha^2 T^4/m^4$), is sufficiently smaller than the second term ($~\alpha^2 T^2/m^2$), and Masood's abc functions ($a_i(m\beta)$)  can be evaluated for the extreme values. The leading order contribution at low T is based on the perturbative expansion in QED and the first term should be proportional to $T^2/m^2$ in this expansion. The damping factor $exp(-m/T)$ appears in the effective action when all the contributing terms are included simultaneously. We look at all these terms one by one and work for sufficiently small range of temperature (for a few MeV) to get the simple and approximate results to evaluate some physically measurable parameters. We ignore the magnetic field effect in this calculation. It is reasonable to demonstrate the renormalizability of QED at low temperature. High densities and magnetic field, contributing to the potential energy, change the scenario [30-34] and the plasmon effect may be studied at higher temperatures but higher order effects with high magnetic field are still to be investigated. 

\bigskip
A comparison of the one loop and two loops contributions of the transverse component of the photon frequency is shown in Figure 3. It is clear that the second order contribution (broken line) is very small as compared to the first order (solid line) contribution. 

\bigskip
\begin{figure}[!hth]
\begin{center}
  \begin{tabular}{c}
    \mbox{\includegraphics[width=4in,angle=0]
{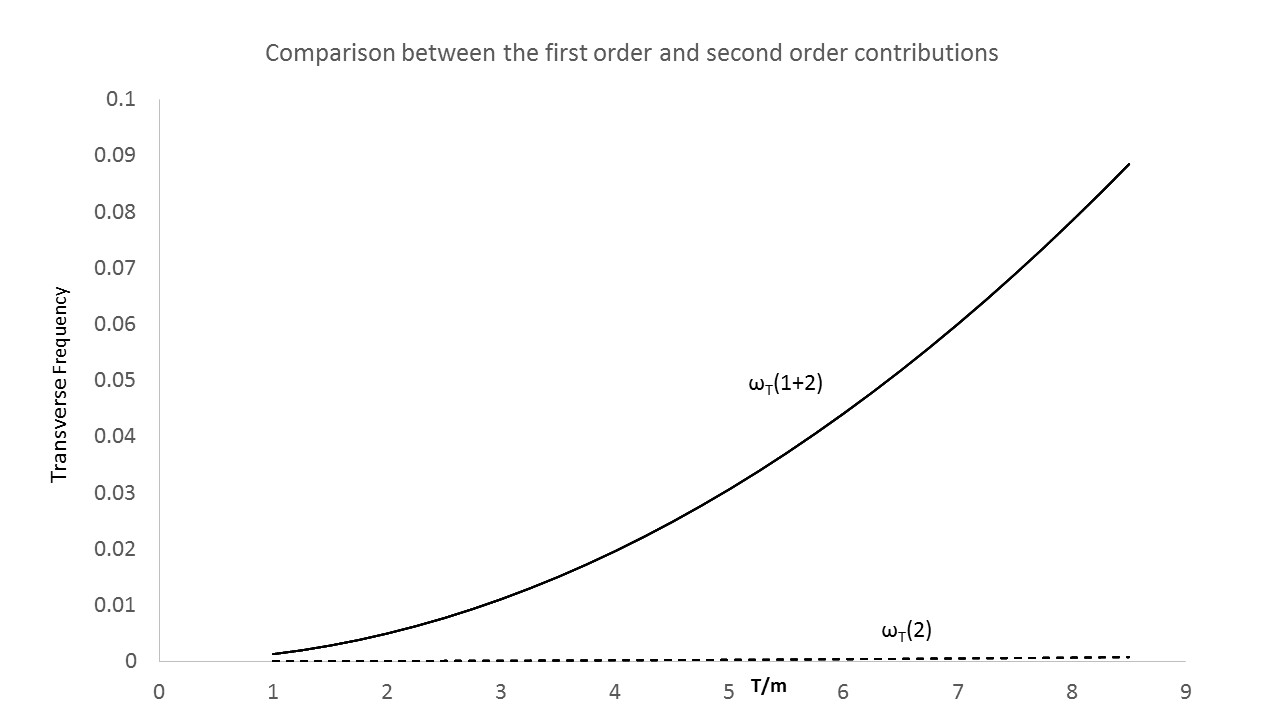}}
  \end{tabular}
\end{center}
\caption{Comparison between the first order and second order contributions to the square of photon transverse frequency $\omega_T^2$.} 
\end{figure} 

A similar behavior is seen between the first order and second order contribution of the longitudinal component of the wavenumber and is shown in Figure 4. It is clear that the second order contribution (broken line) is significantly small as compared to the first order (solid line) contribution. It is evident from Figs, (3) and (4) that the wavenumber of photon gets a much larger thermal contribution than the frequency. However, everything remains smaller than 1 up to the temperatures of 4MeV or so, and this temperature is expected a little before nucleosynthesis in the early universe. 

\begin{figure}[!hth]
\begin{center}
  \begin{tabular}{c}
    \mbox{\includegraphics[width=4in,angle=0]
{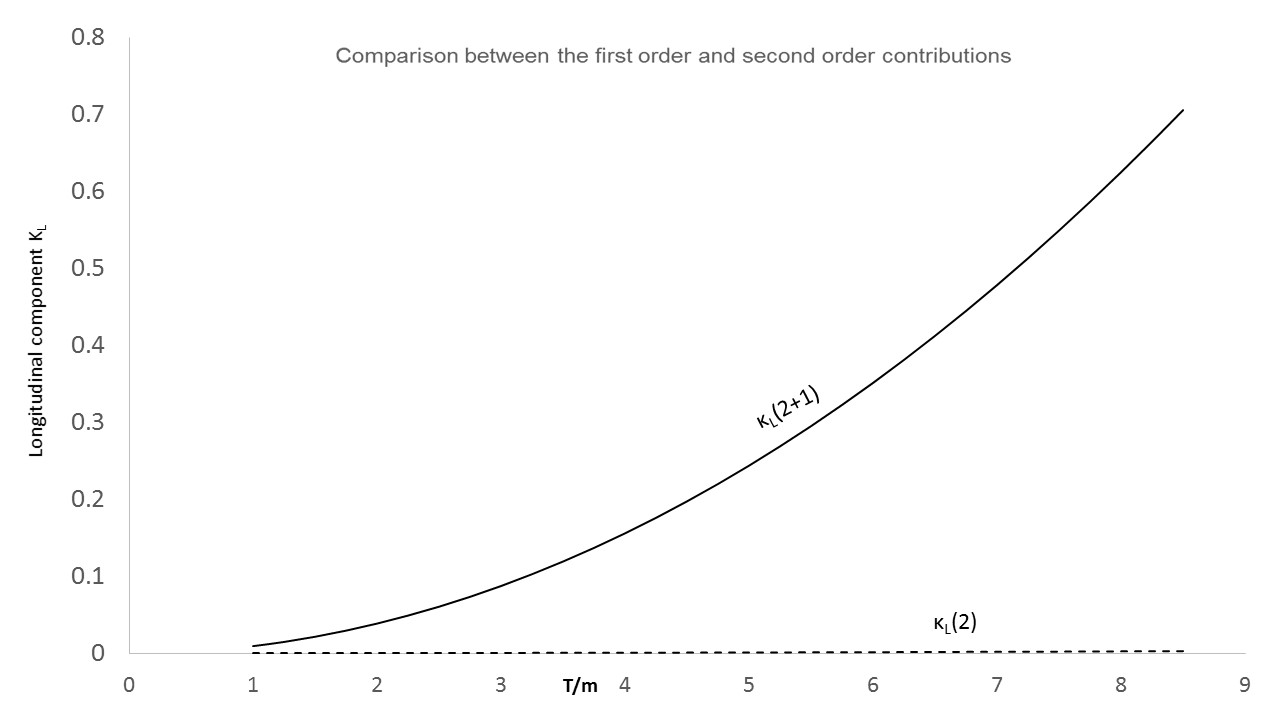}}
  \end{tabular}
\end{center}
\caption{Comparison between the first order and second order contributions to the square of the longitudinal component of the wavenumber $\kappa_L^2$.} 
\end{figure} 

A comparison between Figs.3 and 4 shows that the thermal contributions to $\kappa _L^2$ are smaller than the $\omega _T^2$. We compare the two loop contributions of $\kappa _L$ and $\omega _T$ in Fig. 5. 

\begin{figure}[!hth]
\begin{center}
  \begin{tabular}{c}
    \mbox{\includegraphics[width=4in,angle=0]
{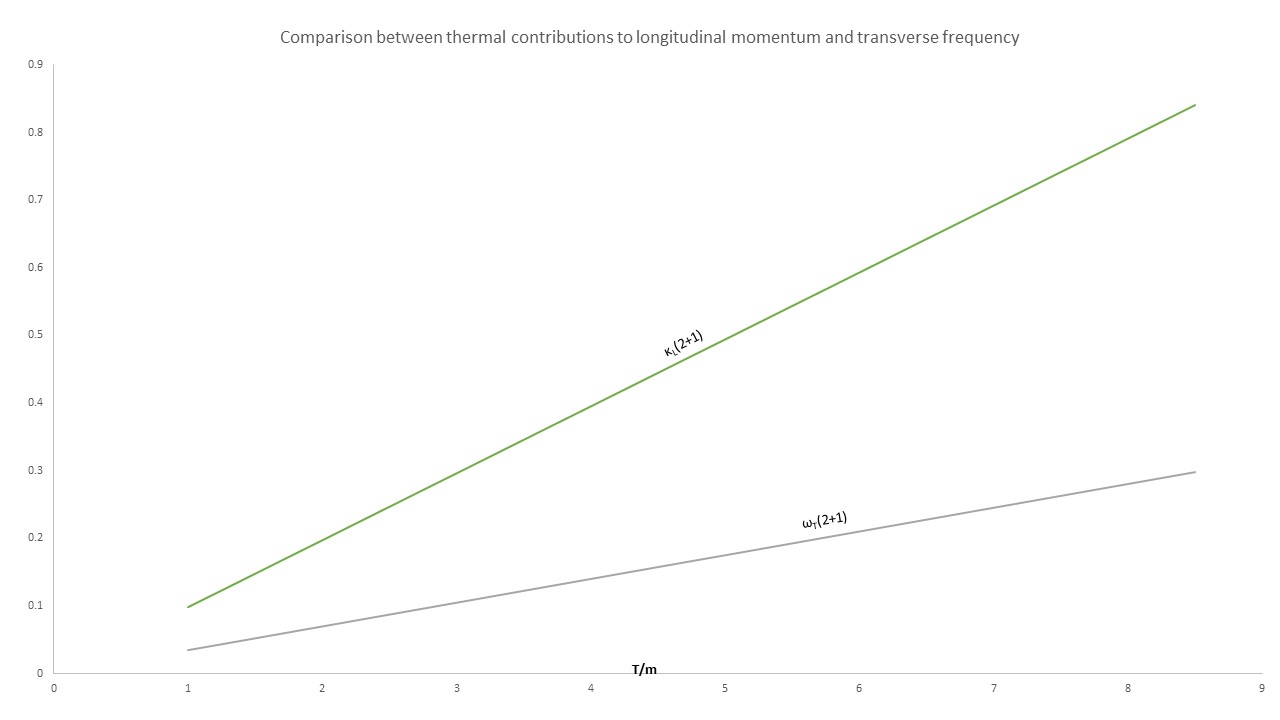}}
  \end{tabular}
\end{center}
\caption{Comparison between the first order and second order contributions to $\kappa _L$ and $\omega _T$.} 
\end{figure} 

It has been shown that the QED coupling behaves the same way in Fig.6. The one-loop thermal correction is much more dominant as compared to the second order contribution which shows the validity of these calculations to prove the renormalizability of the theory. Moreover it indicates the validity of the calculational scheme in this range. Fig.6 shows a comparison between two loop contributions with the corresponding second order contributions  of the charge renormalization constant. However, the  temperature dependent contribution is smaller than other parameters.

\bigskip
\begin{figure}[!hth]
\begin{center}
  \begin{tabular}{c}
    \mbox{\includegraphics[width=4in,angle=0]
{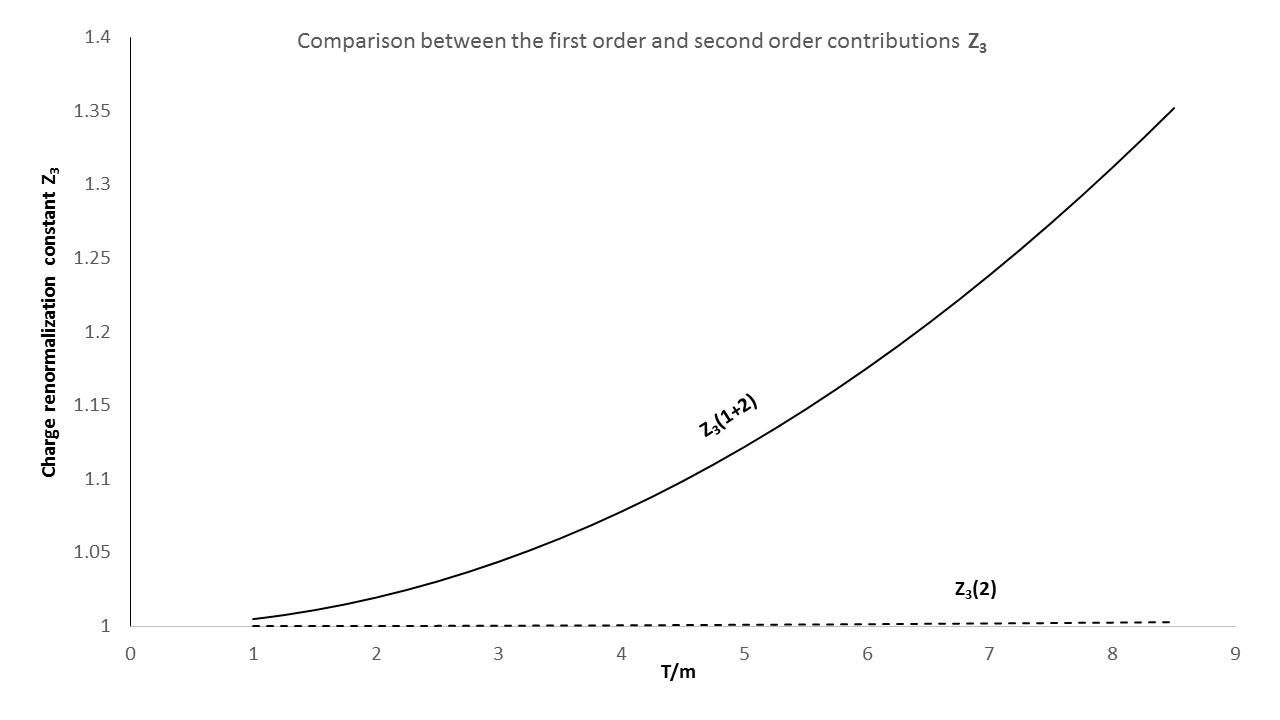}}
  \end{tabular}
\end{center}
\caption{Comparison between the first order and second order contributions.} 
\end{figure}
 
Another interesting outcome of these calculations is Debye shielding in QED plasma, which is plotted as a function of temperature in Fig.7. A comparison between the first order and the second order in $\alpha$ terms clearly demonstrates (Fig.7) that the smaller values correspond to the first order term.It indicates the range of temperatures where plasma phase can exist in QED. 

\bigskip
\begin{figure}[!hth]
\begin{center}
  \begin{tabular}{c}
    \mbox{\includegraphics[width=6in,angle=0]
{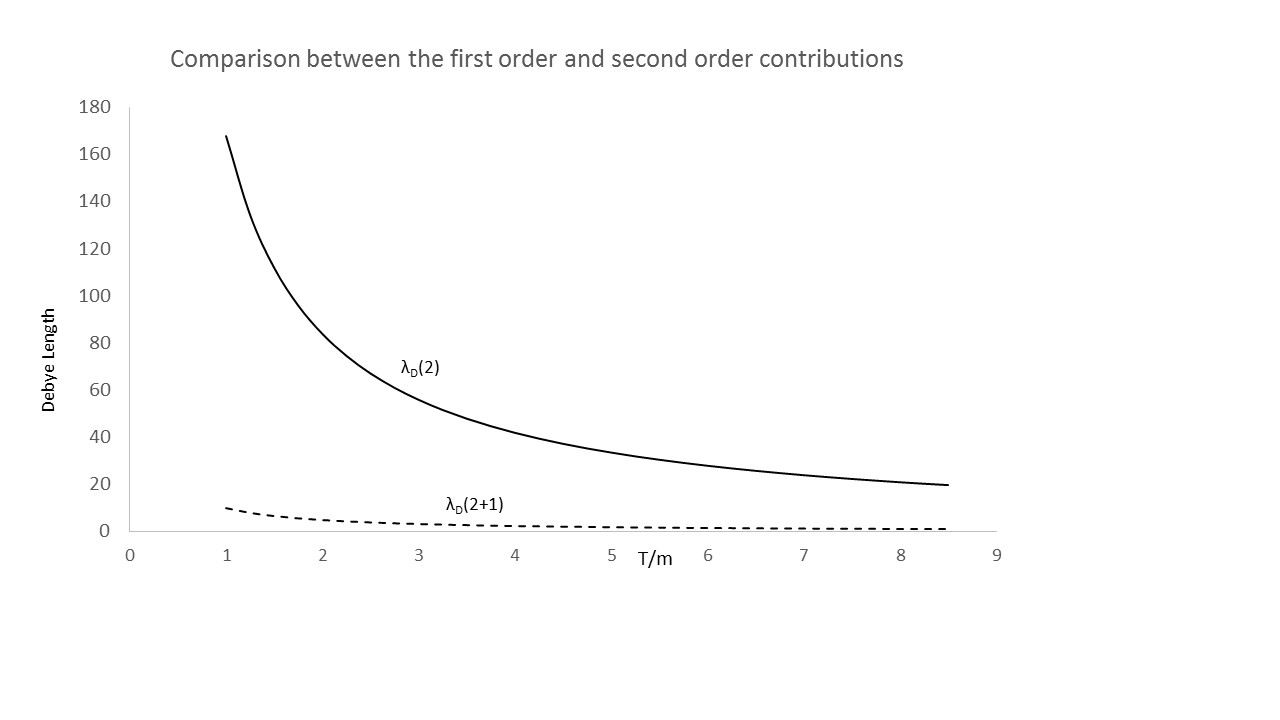}}
  \end{tabular}
\end{center}
\caption{Comparison between the first order and second order contributions.} 
\end{figure} 

In this paper, the first two orders and the second order contributions are compared to validate the renormalizability of QED between $m_e\leq T\leq10m_e$. For this purpose, we plotted the charge renormalization parameter of QED comparing the first two order contributions and the second order contribution only. The quadratic temperature dependence of QED parameters dominates in the proposed temperature range and ensures the validity of renormalizability of QED. The vacuum polarization components, squares of longitudinal and transverse components of photon frequency and momentum are all quadratic functions of temperature. However, the first order contributions indicate the existence of plasmon due to the dominant radiative self-coupling of photon in this region of temperature, which is found a  little before nucleosynthesis and existed throughout the nucleosynthesis. Higher temperature behavior breaks this plasma shielding due to high kinetic energy and photons do not have the ability to couple with the medium at lower temperatures. Therefore the plasmon does not exist outside this range of temperature unless the magnetic field or high density contributions [18,33,34] are incorporated.
\bigskip

\section{REFERENCES}

\begin{enumerate}
\item See for example: C. Itzykson and J. B. Zuber, \textit{Quantum Field Theory }(McGraw- Hill Inc., 1990).
\item Schwinger, J. Math. Phys.\textbf 2 (1961) 407; J. Schwinger, Lecture Notes of Brandeis University Summer Institute (1960).
\item Joseph I.Kapusta and Charles Gale, \textit{Finite Temperature Field Theory Principles and AApplications}( Cambridge Monographs on Mathematical Physics, 1994).
\item S. Weinberg  Phys. Rev.\textbf{ D9}, (1974)3357. 
\item L. Dolan and R. Jackiw, Phys. Rev. \textbf{D9}, (1974) 3320. 
\item  C. Bernard, Phys. Rev. \textbf{D9},  (1974)3312.
\item J. F. Donoghue and B. R. Holstein, Phys. Rev. \textbf{D28}, (1983)340; \textbf{D29}, (1983)3004(E) 
\item P. Landsman and Ch G. Weert, Phys. Rep. \textbf{145},(1987)141  and the references therein.
\item R. Kobes, Phys.Rev.\textbf{D42}, (1990) 562. 
 \item W.Dittrich, Phys.Rev.\textbf{ D19}, 2385 (1979).
 \item P.H. Cox, W.S. Hellman and A.Yildiz, Ann. Phys. \textbf{154}, 211 (1984).
\item T. Kinoshita, J. Math. Phys. \textbf3 (1962) 650; T. D. Lee and M. Nauenberg, Phys. Rev. \textbf{133} (1964) 1549. 
\item A.Weldon, Phys. Rev.\textbf{D26}, (1982)1394.
\item A.Weldon, Phys. Rev.\textbf{D26}, (1982)2789. ; 
\item J. F. Donoghue, B. R. Holstein, and R. W. Robinett, Ann. Phys. (N.Y.)\textbf{164}, (1985)233.
\item K. Ahmed and Samina S. Masood, Ann. Phys. (N.Y.) \textbf{207}, (1991) 460.
\item Samina S.Masood, Euro. Phys. J. C (2017) 77: 826. https://doi.org/10.1140/epjc/s10052-017-5398-0
\item Samina Masood and Iram Saleem, IJMPA. (2017) 32: 17500816.
\item  H. Gies Phys. Rev.\textbf{ D61} (2000) 085021,  hep-ph/9909500).
\item K. Ahmed and Samina Saleem (Masood), Phys. Rev. \textbf{D35}, (1987)1861.
\item K. Ahmed and Samina Saleem (Masood), Phys. Rev. \textbf{D35}, (1987)4020.
\item Samina Saleem (Masood),\ Phys. Rev. \textbf{D36},  (1987)2602.
\item Mahnaz Qader (Haseeb), Samina S. Masood, and K. Ahmed, Phys. Rev. \textbf{D44},  (1991)3322. 
\item Mahnaz Qader (Haseeb), Samina S. Masood, and K. Ahmed,,\textit{\ }Phys. Rev. \textbf{D46}, (1992)5633.
\item Mahnaz Haseeb and Samina S. Masood, Phys. Lett. \textbf{704},(2011)66.
\item Mahnaz Haseeb and Samina S. Masood, IJMPA \textbf{30},(2015)1550198.
\item Duane A.Dicus, David Dawn and Edward W.Kolb, Nucl. Phys.\textbf{B223}, 525(1981).
\item G. A. Hajj and P. M. Stevenson, \textbf{37}. (1988) 413.
\item Samina Masood, Phys.Res.Int 2014, 48913(2014)1 (arXiv:1407.1414)
\item See for example Ref.[8,9], Samina S. Masood, Phys. Rev. \textbf{D48}, (1993)3250;
\item Samina S.Masood, Astroparticle Physics \textbf{4}, (1995)189 .
\item Samina Masood,JHEPGC , Vol.01 No.01(2015), Article ID:56270,12 (arXiv:1506.01284 [hep-ph])
\item Samina S. Masood, Phys. Rev. \textbf{D44}, (1991)3943.  
\item Samina S.Masood, Phys. Rev. \textbf{D47}, (1993)648. 
\end{enumerate}
\end{document}